# The breakdown of the direct relation between the density scaling exponent and the intermolecular interaction potential for molecular systems with purely repulsive intermolecular forces.

F. Kaśkosz[1,2*], K. Koperwas[1,2], A. Grzybowski[1,2] and M. Paluch[1,2]

1. University of Silesia in Katowice, Institute of Physics, 75 Pułku Piechoty 1, 41-500 Chorzów, Poland
2. Silesian Center for Education and Interdisciplinary Research SMCEBI, 75 Pułku Piechoty 1a, 41-500 Chorzów, Poland

* corresponding author: filip.kaskosz@us.edu.pl

## ABSTRACT

In this work, we question the generally accepted statement that the character of intermolecular interactions can be directly determined from the scaling exponent. Based on detailed studies of polyatomic molecular systems with precisely defined and purely repulsive intermolecular potential, we show that the value of the density scaling exponent evidently differs from the one predicted by the intermolecular virial-potential-energy correlation. Since the latter value directly results from the intermolecular potential, information on the interactions between molecules within the system cannot be immediately gained from the density scaling exponent value. Moreover, we suggest that the recently proposed "molecular force" method also returns the value that varies from the one scaling the dynamics. Finally, basing on our results, it might be deduced that the intramolecular interactions influence the density scaling value for real liquids.

**MANUSCRIPT**

Many liquids have a glass-forming ability that often makes them capable of becoming supercooled, i.e., they may remain a liquid even when their temperature is lowered below the melting point. The supercooling state can be achieved by isobaric cooling, where a constant pressure ensures that the melting point is fixed, or the liquid can be isothermally compressed. In the second scenario, transformation from normal liquid to metastable state results from an increase in pressure, which causes the molecules to be more densely packed. In turn, lowering the temperature is equivalent to reducing the average kinetic energy of the molecules while increasing the pressure forces the molecules to get closer to each other, which decreases the volume in which the molecules can move. As a result of both processes, the liquid dynamics slowdown proceeds in the Arrhenius manner: the dependence of dynamic quantities on temperature (or pressure) is exponential.[1] In the supercooling regime, however, the dynamics slowdown trend changes and becomes non-exponential.[2,3] The critical factor determining the non-trivial nature of liquid dynamics is not fully defined[4,5], but it is recognized that the temperature and volume effects are responsible. Which of them plays a more important role has been the subject of many works and widely discussed.[6] It was found that the temperature and volume relative influence differs among the substances. Moreover, it was suggested[6] that instead of defining the factor with the dominant role, one should rather consider some function of these properties to represent the dynamics properly. A breakthrough that contributed to simplifying the description of liquid dynamics is the density scaling concept.[7,8] According to this concept, dynamics under different thermodynamic conditions can be mapped into a single relationship. The dynamical quantities can be expressed using the function of a single variable, $TV^\gamma$, the product of temperature and volume raised to the power of the scaling exponent, from which inferences about the relative influence of temperature/volume on molecular dynamics can be made.[6,9] Experimental studies[10] performed for more than 100 real glass-forming liquids showed that the value of the scaling exponent varies for different substances, indicating that $\gamma$ is a material parameter. A problem about what determines the value of the scaling exponent of molecular systems arose. Based on the theoretical research[11,12] and the results of the computer simulations[13,14] of the system described by the inverse power-law interaction potential, the physical interpretation of the scaling exponent was suggested. For the simple monoatomic system whose interactions were described using a potential in the form of an inverse power law[11] $U_{IPL}(r) = Cr^{-m} + A$ ($C$ - potential parameters, $A$ – attractive background constant, $r$ – distance), the scaling exponent turned out to be related to the potential as follows: $\gamma = m/3$.

By analogy, the explanation of the scaling exponent value of the monoatomic system governed by the Lennard-Jones interaction potential $U_{LJ}(r) = 4\epsilon \left(\frac{\sigma}{r}\right)^{12} - 4\epsilon \left(\frac{\sigma}{r}\right)^{6}$, ($\epsilon$ - the potential well depth, $\sigma$ - the distance at which the potential energy is zero) was made.[15] At short distances where the potential is purely repulsive, up to the characteristic radius where it has a minimum, $U_{LJ}(r \leq 2^{1/6}\sigma)$ was fitted by a single effective $U_{IPL}$. The fit takes into account that the potential's attractive term modifies the curvature of the repulsive term. Consequently, the effective $U_{IPL}$ potential describes the repulsion with exponent $m_{eff} \approx 18$, which explains $\gamma_{LJ} \approx 6$. However, at this point, the shortcomings of this consideration are worth mentioning. Only the part corresponding to the purely repulsive potential is taken into account, and this implies the misleading conclusion that repulsive forces exclusively determine the dynamics. If indeed the dynamics are governed only by repulsive forces, then both the considered $LJ$ system and the corresponding system with $LJ$ potential truncated at the minimum and shifted (i.e., Weeks-Chandler-Andersen, $WCA$) should have the same value of scaling exponent. However, it was shown that both the temperature dependences of the relaxation times and the scaling exponents of the systems are different from each other[16–18], which emphasizes the importance of the role of attraction, which cannot be ignored. The second issue is that the effective potential method works only for the systems characterized by a potential distribution with spherical symmetry. In the case of real systems, it should be considered that molecules commonly consist of many atoms bound into different structures. Additionally, but not necessarily, these atoms can be of various types. Due to these reasons, the potential around the real molecules is non-uniformly distributed in different directions. Therefore, anisotropy is an inherent feature of real molecular systems that must be taken into account. In connection with this, it is highly challenging to correctly represent even the simplest anisotropic (i.e., diatomic) system by a point particle. The task requires specifying the reference point where the potential should have its source and considering that the value it takes depends not only on the distance but also on the orientation relative to some specific directions. It is also important that when considering the interaction of two molecules, each atom of the first one affects each atom of the second one according to a given potential, due to which it exerts forces in specific directions. Thus, the forces resulting from the impact of the first molecule on individual atoms of the second one depend on the mutual arrangement of both molecules. Therefore, the effective potential must correctly reflect the net forces between molecules.

A promising and systematically developed approach that seems to overcome the aforementioned problems is the isomorph theory[19–23], according to which the scaling occurrence results from a strong correlation between potential energy $U$ and the configurational contribution to pressure virial $W$. In the case of the monoatomic $IPL$ system, the simplicity of the potential implies that the $W - U$ correlation is perfect, so its relationship is linear, and its slope coefficient is equal to $m/3$. Considering more complex potentials whose curvature due to the attractive term varies with distance, such as $LJ$, the correlation is weakened. However, the theory states that it is sufficient for the correlation coefficient to be greater than or equal to 0.9, and then for such highly correlated systems (Roskilde-simple liquids), the scaling exponent can be determined as the slope of the $W - U$ dependence. In this way, the scaling exponent was successfully obtained for, among others, the monoatomic $LJ$, the Kob-Andersen binary $LJ$ mixture, and some molecules with rigid bonds.[24–26] Considering the rigid dumbbells example, it can be concluded that the correlation slope allows the determination of a scaling exponent for anisotropic systems. On the other hand, the results obtained for the monoatomic Gay-Berne system of ellipsoidal-like molecules contradict this statement.[27] Similarly, the study of quasi-realistic systems[28], i.e., rhombus-like $LJ$ molecules with flexible bonds, $RM$, demonstrated that the scaling exponent cannot be determined from the dependency of the total $W$ and $U$ quantities because the contribution from their intramolecular interactions' terms significantly impairs the correlation.[29] Therefore, the focus was on the virial and potential energy terms derived from intermolecular interactions, for which high correlation is present. Nevertheless, the slope of their relationship does not provide a scaling exponent. It was suspected that the potential curvature (an exponent of the effective $IPL$ potential that could match the $LJ$ potential within narrow intervals at given distances) changing with distance was responsible for this. Hence, in order to eliminate this factor, the consequent studies focused on the simplified system $RM_{IPL}$, which differs from $RM$ in the way that all intermolecular interactions were described by the purely repulsive potential, $U_{IPL}$ with exponent $m = 12$. The dynamical quantities of $RM_{IPL}$ were successfully scaled with an exponent equal to 4, i.e., with the value of the intermolecular virial and potential energy correlation slope. In the same work, a tetrahedral-shaped $IPL$ system ($TM_{IPL}$) was also investigated. Surprisingly, the density scaling exponent for the $TM_{IPL}$ system was also found to be equal to 4, which suggested that in the case of $IPL$ molecules, the value of the scaling exponent may not depend on the molecule's structure. This finding initiated further studies[30] of the $IPL$ molecular systems with different potentials' parameters, as a result of which, an exact definition of the scaling exponent for polyatomic $IPL$ molecules was obtained: $\gamma$ is the weighted average of the exponents of potentials describing the particular intermolecular interactions, where the weights are the potential energies corresponding to these interactions.

In this work, we challenge the generally accepted statement that the scaling exponent for molecular systems can be identified with the exponent of the potential describing inter-molecular interactions. We conduct comprehensive research on the molecular systems that lack intermolecular attractive forces, investigating whether the scaling exponent is related to the $W - U$ dependency. Moreover, we check whether, for $IPL$ systems with arbitrary molecular structure, the scaling exponent takes the value of the exponent describing the $IPL$ potential. Additionally, in the final stage of research, we apply the "molecular force" method, a recently developed tool based on isomorph theory, to trace the invariant dynamics and structure's phase-diagram curves to check if we can use it to determine the scaling exponent.

The systems we study in this work are constructed on the basis of $RM_{IPL}$ and $TM_{IPL}$[31–34], with one additional atom attached to their structures, from now on referred to as $RM - tail_{IPL}$ and $TM - tail_{IPL}$. In the case of $RM - tail_{IPL}$, the additional atom is attached to the atom located at the end of the longer diagonal of the rhombus, and for $TM - tail_{IPL}$, the extra atom is bonded to one of the side atoms. The schemes of the molecular structures can be seen in the insets of FIG. 1.

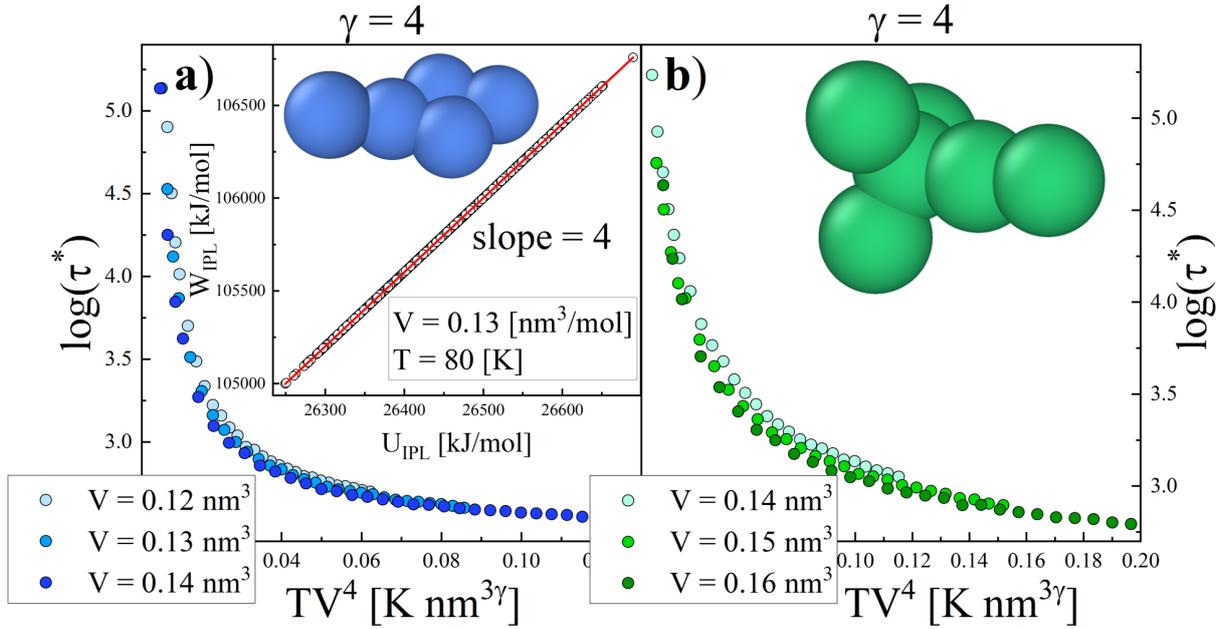

FIG. 1 Reduced relaxation times for mass centers of molecules as a function of the density scaling variable with an exponent determined from the correlation of the virial and potential energy corresponding to intermolecular interactions, see inset in a) section. Schematic representation of the molecular structure of $RM - tail_{IPL}$ on the left and $TM - tail_{IPL}$ on the right.

All constants describing intra-molecular interactions (stiffness of bonds, angles, and dihedrals) and inter-molecular interactions were set in correspondence to the Optimized Potentials for Liquid Simulations[35] ($OPLS$) all-atom force field for carbon atoms in the benzene ring. For $RM - tail_{IPL}$: all bonds have length $r_{B0}^{RM} = 0.14982 nm$, the molecule has three equilibrium valence angles $\theta_0^{RM} = [53.13°, 126.87°, 153.435°]$, and a dihedral described by the Ryckaert–Bellemans function with constants $C^{RB} = [30.334, 0.0, -30.334, 0.0, 0.0, 0.0] \frac{kJ}{mol}$. Therefore, the atom attached as a tail can make up-down movements. For $TM - tail_{IPL}$: bonds have the length $r_{B0}^{TM} = 0.14 nm$, and there are two equilibrium valence angles $\theta_0^{TM} = [109.5°, 180°]$. The stiffnesses of the bonds and angles for both molecules are described by stretching constants: $k^{bond} = 392459.2 \frac{kJ}{mol \cdot nm^2}$, $k^{angle} = 527.184 \frac{kJ}{mol \cdot rad^2}$. Molecules are composed of identical atoms with the following parameters: $\epsilon = 0.2928 \frac{kJ}{mol}$, $\sigma = 0.355 nm$, and their nonbonded, intermolecular interaction is described by the $U_{IPL}$ potential with exponent $m = 12$, where $C = 4\epsilon\sigma^m$. The interaction potential $U_{IPL}$ is truncated at a distance $r_{cut} = 1.065 nm$ (which corresponds to a triple value of the $\sigma$ parameter) and shifted by the constant $A = -U_{IPL}(r_{cut})$ to ensure it is equal to zero at the cut-off distance.

In the first stage of the research, we performed molecular dynamics simulations using the GROMACS software.[36–38] Systems consisting of 4000 molecules were examined in the pressure range from 0 to 100 MPa, customary for standard experiments. The procedure included cooling under isochoric conditions of $V = 0.12, 0.13, 0.14 \frac{nm^3}{mol}$ (for $RM-tail_{IPL}$) and $V = 0.14, 0.15, 0.16 \frac{nm^3}{mol}$ (for $TM-tail_{IPL}$) with a $10\ K$ temperature decrement, using a Nose-Hoover thermostat.[39–41] The $NVT$ simulations for a single thermodynamic point consisted of two parts, the first of which was devoted to the equilibration of the system lasting at least ten times longer than the relaxation time $\tau$ determined from the incoherent intermediate scattering function ($IISF$)[42] calculated for the molecular centers of mass using the wave vector corresponding to the maximum of the structure factor. In the second part, a simulation was performed during which data was collected. The relaxation time $\tau$ was estimated as the time after the $IISF$ function takes the value $1/e$. Subsequently, because the isomorph theory assumes that scaling can be observed when the quantities of the system are considered in reduced units, the relaxation times were reduced according to the formula $\tau^* = \frac{\tau}{V^{1/3}\sqrt{m_{mol}/(k_B T)}}$, where $k_B$ is the Boltzmann constant and $m_{mol}$ is the mass of the molecule (the mass of the atom $m_{at} = 12.011\ u$).[22]

Then, due to the lack of correlation between total virial and potential energy, we calculated the $W, U$ contributions resulting from intermolecular interactions.[29,30,43] Because of the simplicity of the $IPL$ potential, the relationship is perfectly linear and has a slope coefficient equal to $4$, see inset of FIG. 1a). The main parts of FIG. 1 show the relationship between the reduced relaxation times and the density scaling variable with exponent $\gamma = 4$, which does not collapse well onto one curve. Hence, the slope of the correlation does not give a scaling exponent. To check whether the relaxation times of the system can be scaled into one curve, we used the density scaling criterion that $(TV^\gamma = const)_{\tau^*=const}$ and so $\gamma = -\left(\frac{d(\log_{10} T)}{d(\log_{10} V)}\right)_{\tau^*=const}$. Based on the Vogel-Fulcher-Tammann[44–46] ($VFT$) function fit to the temperature dependence of the reduced relaxation times, we determined the temperature and volume conditions in which $\tau^*$ is constant ($\tau^* = 3,\ 3.5,\ 4$), see insets of FIG. 2. As a result, we obtained linear volume-temperature relationships with the slopes equal to: $4.87 \pm 0.03$ for $RM-tail_{IPL}$, and $5.07 \pm 0.22$ for $TM-tail_{IPL}$. Consequently, in the main parts of FIG. 2, we confirmed that the reduced relaxation times can be scaled into a single curve with the exponent equal to the slope obtained from the density scaling criterion. Unless a slight deviation for long relaxation times may be observed for $TM-tail_{IPL}$.

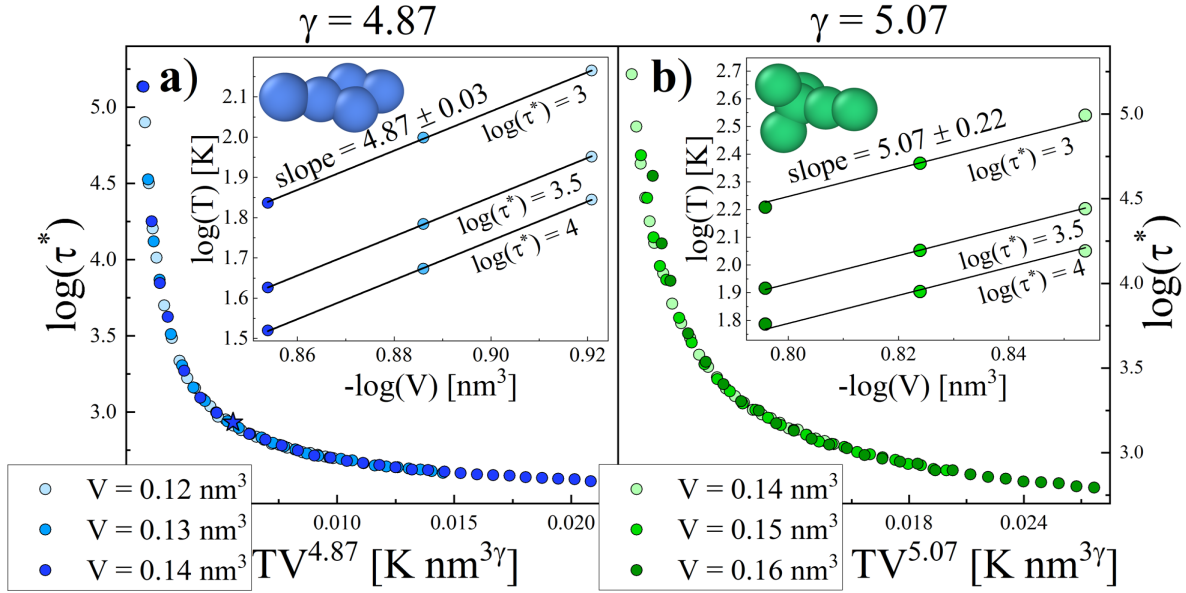

FIG. 2 Reduced relaxation times for mass centers of molecules as a function of the density scaling variable with an exponent determined from the density scaling criterion, see insets. $RM-tail_{IPL}$ in the a) section and $TM-tail_{IPL}$ in the b) section. The point marked with the star symbol is used in testing the molecular force method.

The above result confirms that the scaling exponent for $IPL$ molecular systems with only one type of inter-molecular interactions may have a different value than the exponent of the $IPL$ potential. The discrepancy between the scaling exponent and the $IPL$ potential exponent confirms that the intermolecular interaction potential is not the only factor determining the scaling exponent for molecular systems. A possible explanation is that the scaling exponent is structure-dependent. Therefore, based on the scaling exponent, we cannot directly infer the characteristics of the intermolecular interaction potential, even for simplified molecular systems, i.e., with purely repulsive intermolecular interactions.

The molecular system described by the $IPL$ potential for which the scaling exponent was obtained from the $W-U$ correlation is a rigid dumbbell.[24] In our work, inspired by the above example, we check whether the $W-U$ relationship will allow us to determine the scaling exponent when we consider an analogous system but with a flexible bond. Appropriately, we designed the $DB_{IPL}$ symmetric dumb-bell system, which are described by the same parameters of inter-molecular interactions as tailed systems. The only intra-molecular interaction is a harmonic bond between two identical atoms characterized by the same length, and the same bond stretching constant as the bond for the tetrahedral system. To consider the analogous pressure range, we performed the simulations for the system consisting of 5000 molecules in constant volume conditions $V = 0.06, 0.07, 0.08 \frac{nm^3}{mol}$. The simulation procedure and the analysis of the obtained data were identical, as in the case of systems with tails. Using the dependency of relaxation times calculated from $IISF$, we determined a scaling exponent for $DB_{IPL}$ equal to 3.09, see FIG. 3. For real molecules, $\gamma < 4$ was observed for extremely polar systems.[47] In the case of $DB_{IPL}$, the system is non-polar, yet the scaling exponent takes a value relatively much smaller than 4.

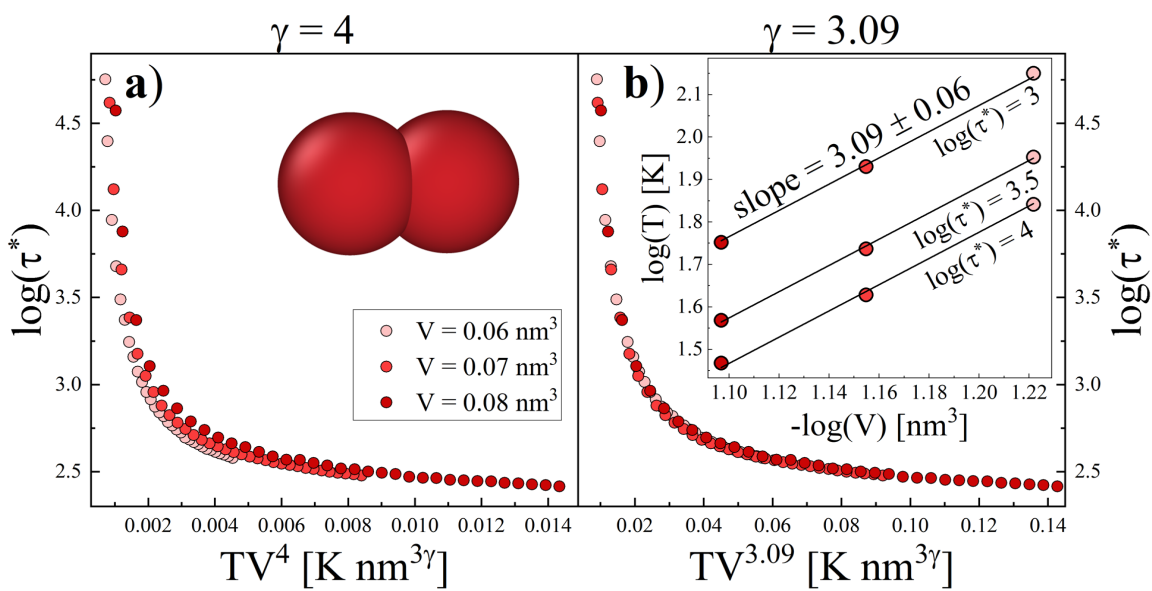

FIG. 3 Reduced relaxation times for mass centers of molecules as a function of the density scaling variable with an exponent determined from the correlation of the virial and potential energy corresponding to intermolecular interactions on the left and from the density scaling criterion on the right. The inset of section a) shows the scheme of the $DB_{IPL}$ molecular structure, and the inset of section b) shows the density scaling criterion.

In FIG. 3, it can be seen that the relaxation times collapse well to one curve when using the exponent determined from the scaling criterion while the $W - U$ correlation does not provide a scaling exponent. Similar result is presented in the work on the system of molecules of Lennard Jones chains with flexible bonds.[43,48] It was suggested that for systems with harmonic bonds the phase diagram curves with invariant structure and dynamics be called pseudoisomorphs. This is because the invariance for molecular systems is approximate. Namely, it can be found for the dynamics and structure of the centers of mass, but it is highly impaired for the atomic structure and dynamics, especially for systems with flexible bonds. Due to the inability to use the virial and potential energy relationships to determine the scaling exponent of molecular systems, a search for new ways of tracking invariance began.

One of the recently developed approaches is the so-called "force method"[25,49], according to which, based on a single configuration obtained from a system simulation, it is possible to find isomorphic thermodynamic points. Briefly speaking, let's consider a starting configuration containing information about the atoms' position in the system. Its isotropic rescaling allows for obtaining a target configuration whose structure is enlarged or shrunk in the same way in all directions. In reduced units, both configurations are structurally identical. According to the theory, for both states (starting and target) to be isomorphic, the reduced forces exerted by the system on its atoms in both states should also be the same. In consequence, considering a system with the density $\rho_{init}$ in the starting configuration characterized by the vector $\boldsymbol{R}_{init} = (\boldsymbol{r}_1, \boldsymbol{r}_2, \ldots, \boldsymbol{r}_N)_{init}$ of the $N$ atom's positions, for a given target density $\rho_{target}$, we can create a target configuration $\boldsymbol{R}_{target} = \left(\frac{\rho_{init}}{\rho_{target}}\right)^{1/3} \boldsymbol{R}_{init}$ and estimate the target temperature from the equation $T_{target} = \frac{|F(\boldsymbol{R}_{target})|}{|F(\boldsymbol{R}_{init})|} \left(\frac{\rho_{init}}{\rho_{target}}\right)^{1/3} T_{init}$, where $|\boldsymbol{F}(\boldsymbol{R})|$ is a length of the vector of forces exerted by the system on its atoms. As a result, the obtained states of the system at the initial and target conditions should be isomorphic. Consequently, on their basis, the scaling exponent can be determined from the formula $\left(\frac{\rho_{target}}{\rho_{init}}\right)^{\gamma+1/3} = \frac{|F(\boldsymbol{R}_{target})|}{|F(\boldsymbol{R}_{init})|}$. This method has been successfully applied to atomic systems and systems of rigid molecules.[25,49]

Now, we check whether this approach can be successfully applied to molecules with flexible bonds. The method will be tested for our system $RM - tail_{IPL}$. As the starting point, we choose the thermodynamic point $V_{init} = 0.14 \frac{nm^3}{mol}$, $T_{init} = 80K$, and as the target state $V_{target} = 0.12 \frac{nm^3}{mol}$ that is about 17% more than the starting density. The starting point used in testing the molecular force method is marked in FIG. 2a) with the star symbol. In the case of molecular systems, configuration rescaling cannot be made at the atomic level because it would change the bond lengths. Due to that, the arrangement of individual molecules is left the same as in the starting configuration, and only the positions of the centers of mass are rescaled. To take statistics into account, we analyzed 200 configurations from the equilibrated system simulation. In the molecular systems case, only forces from non-bonded interactions of surrounding molecules contribute because the forces resulting from intra-molecular interactions reduce. Hence, the method does not take into account intra-molecular interactions. Using the formula $T_{target} = \frac{|F(\boldsymbol{R}_{target})|}{|F(\boldsymbol{R}_{init})|} \left(\frac{\rho_{init}}{\rho_{target}}\right)^{1/3} T_{init}$, we determined

temperature in the target state $T_{target} = 196.58\ K$. Consequently, we simulated the system at the target thermodynamic point to verify the system's characteristics directly in reference to the starting state. In the FIG. 4, we compare the dynamics via $IISF$ for the centers of mass as a function of reduced time for $RM - tail_{IPL}$.

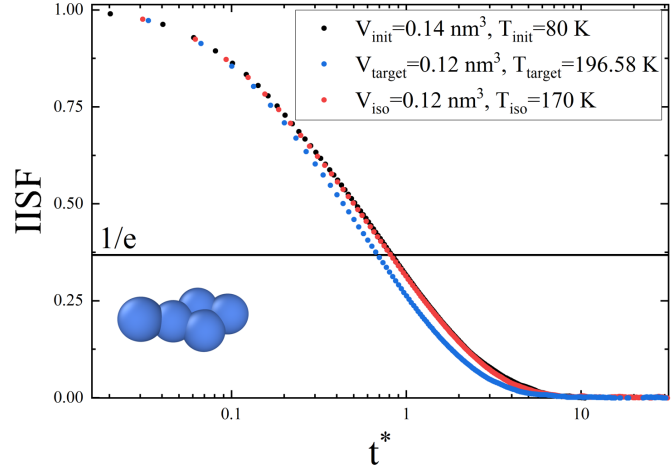

FIG. 4 Incoherent intermediate scattering function for centers of mass in function of the reduced time calculated for $RM - tail_{IPL}$ at points: starting, target, and determined from the density scaling criterion. The relaxation time is determined as the time the IISF function gets the value $1/e$.

FIG. 4 shows that the functions are different, so the starting and the target points are not characterized by invariant dynamics. The faster decay of the $IISF$ function calculated for the target point confirms that the temperature determined using the molecular force method is overestimated. To quantify the overestimation, we placed the $IISF$ function in the figure for one more thermodynamic point, which has invariant dynamics with the starting point and was determined based on the scaling exponent ($T_{iso} = T_{init} \left(\frac{V_{init}}{V_{target}}\right)^{\gamma=4.87} = 169.48 \approx 170K$). As a result, the overestimation of the temperature estimated from the molecular force method for our system is at least $26\ K$.

Summarizing, this work focused on verifying the generally accepted claim that the density scaling exponent is directly related to the intermolecular potential. Within it, we carried out detailed studies of molecular systems characterized by harmonic intramolecular interactions and no intermolecular attraction. We checked that the methods using the interaction potential, i.e., virial-potential energy correlation and molecular force, did not enable us to determine the scaling exponent for our systems. Therefore, relying solely on the intermolecular potential is insufficient to determine the scaling exponent of the molecular systems, and that, in turn, highlights the possible importance of their structural anisotropy and intramolecular interactions. In conclusion, it must be emphasized that based on the scaling exponent's value, we cannot directly infer the characteristics of the intermolecular potential of real liquids.

**ACKNOWLEDGEMENTS**

The author thanks Andrzej Grzybowski for supportive discussions and guidance. The authors are deeply grateful for the financial support by the National Science Centre of Poland within the framework of the Maestro10 project (Grant No. UMO-2018/30/A/ST3/00323).